\documentclass[showpacs,amssymb,aps]{revtex4}
\usepackage{graphicx}

\begin{document}

\title{Light-front field theories at finite temperature}

\author{V. S. Alves}\altaffiliation{Permanent address: Departamento
de  F\'{\i}sica, Universidade
Federal do Par\'{a}, 66075-110 Bel\'{e}m, Brasil}
\author{Ashok Das}
\author{Silvana Perez}\altaffiliation{Permanent address: Departamento
de  F\'{\i}sica, Universidade
Federal do Par\'{a}, 66075-110 Bel\'{e}m, Brasil}
\affiliation{Department of Physics and Astronomy,
University of Rochester,
Rochester, NY 14627-0171, USA}

\bigskip

\begin{abstract}

We study the question of generalizing light-front field theories to
finite temperature. We show that the naive generalization has serious
problems and we identify the source of the difficulty. We provide a
proper generalization of these theories to finite temperature based on
a relativistic description of thermal field theories, both in the real
and the imaginary time formalisms. Various issues associated with
scalar and fermion theories, such as
non-analyticity of self-energy, tensor decomposition are discussed in detail.
\end{abstract}

\pacs{11.10.Wx, 11.10.Kk, 12.38.Lg}

\maketitle

\section{Introduction}

Quantum field theories are conventionally quantized on a
space-like surface. However, quantization on a light-like surface
provides an interesting alternative which dates back to the works of
Dirac \cite{dirac}. Light-front field theories \cite{chang1,chang2},
namely theories quantized on a 
light-front, have found applications in various branches of
physics such as QCD, string theories and membrane theories, among
others \cite{reviews}. In the case of QCD, for example, they provide a
method for 
studying non-perturbative phenomena systematically, which is
different from the usual lattice studies.

In fact, there is a lot of activity in the study of light-front
field  theories within the context of QCD. Many of the features of
light-front theories are quite distinct from the conventional
equal-time field theories - one of the most significant being that
light-front theories are first order (as opposed to conventional
theories which are second order) in time derivatives and,
correspondingly, describe different degrees of freedom. While the
behavior of conventional field theories at finite temperature are
quite well understood by now \cite{kapusta,lebellac,das}, a systematic
study  of the thermal
properties of light-front field theories is lacking so far. It is
the purpose of this paper to work out the essential properties of
scalar and fermion light-front field theories at finite
temperature. We will defer the study of light-front gauge theories
to a future publication. 

It would seem that the generalization of
the thermal field concepts to light-front theories should be
straightforward. In fact, in $n$-dimensions, if we define the
light-front time variable as
\begin{equation}
x^{+} = \frac{1}{\sqrt{2}} \left(x^{0} + x^{n-1}\right)
\end{equation}
then, the Hamiltonian of the system can be identified with
\begin{equation}
H = P^{-} = \frac{1}{\sqrt{2}} \left(P^{0}-P^{n-1}\right)
\end{equation}
This would suggest that the ensemble average of an arbitrary operator,
${\cal O}$, in the light-front field theory, should be defined as
\begin{equation}
\langle {\cal O}\rangle_{\beta} = {\rm Tr}\; e^{-\beta H}\, {\cal O} =
{\rm Tr}\; e^{-\beta P^{-}}\,{\cal O}
\end{equation}
where $\beta$ represents the inverse temperature in units where
the Boltzmann constant is unity. This has been, in fact, the
general  thinking \cite{brodsky}. However, as we will show, this
straightforward
generalization is incorrect and leads to various problems. We
will discuss the proper description of thermal field theories (for
scalars and fermions), on the light-front, both in the imaginary as
well as the  real time
formalisms. In section {\bf II}, we will describe the naive
generalization of the concepts of thermal field theories to
light-front  scalar field theories, both in the imaginary and the real
time  formalisms, and show that this leads to various problems. The
source of the difficulty is identified in section {\bf III}, where
we give the proper description of light-front scalar
theories at finite temperature. We calculate various quantities of
interest  such as the
thermal mass correction as well as the non-analyticity in the
self-energy. We show that there are more possible limits that can arise in
these theories, in contrast to conventional thermal field theories. A
short  description of the tensor decomposition, which has a richer
structure 
in such theories, is also given. In section {\bf IV}, we discuss
briefly the generalization of light-front fermion theories to
finite temperature. Here a new feature arises since these theories
have only half the number of independent degrees of freedom
\cite{chang2}. The
generalization of light-front gauge theories to finite temperature
as well as various other applications are under study and will be
described in a future publication.

\section{Naive generalization to finite temperature}

In this section, we will discuss the naive generalization of the
techniques of thermal field theory to the light-front scalar field
theories. Let us briefly establish the notation. In
$n$-dimensions, we define
\begin{equation}
x^{\pm} = \frac{1}{\sqrt{2}} \left(x^{0} \pm x^{n-1}\right)
\end{equation}
where $x^{+}$ is identified with the light-front time coordinate.
Denoting the  coordinate vector as $x^{\mu} =
(x^{+},x^{-},\vec{x})$, where $\vec{x}$ represent the $(n-2)$ transverse
coordinates, it is easy to see that the nontrivial components of
the metric in this basis have the form
\begin{equation}
\eta^{+-} = \eta^{-+} = 1,\qquad \eta^{ij} = - \delta^{ij}
\end{equation}
so that the scalar product of two vectors can be written as
\begin{equation}
A\cdot B = A^{+}B^{-} + A^{-}B^{+} -\vec{A}\cdot \vec{B}
\end{equation}
The momentum vector can also be written as $p^{\mu} =
(p^{+},p^{-},\vec{p})$, where $p^{-}$ can be identified with the
energy variable. In the light-front variables, the Einstein
relation takes a linear form
\begin{equation}
p^{-} = \frac{\vec{p}^{\;2}+m^{2}}{2p^{+}}
\end{equation}
This is a major difference from the conventional quantization on
an equal-time surface.

In the light-front variables, the Lagrangian density for a $\phi^{4}$
theory, for example, takes the form
\begin{equation}
{\cal L} = \partial_{+}\phi \partial_{-}\phi - \frac{1}{2}
(\vec{\nabla}\phi)^{2} - \frac{m^{2}}{2} \phi^{2} -
\frac{\lambda}{4!} \phi^{4}\label{lagrangian}
\end{equation}
It is clear that the Euler-Lagrange equations, following from this,
are only  first order
in the $x^{+}$ derivative, which is a distinctive
feature of light-front theories. The quantization of this theory
has been discussed quite a lot in the literature and without going
into details, we simply note here that the Feynman rules for this
theory, at zero temperature, take the form
{\renewcommand{\arraystretch}{0.5}
\begin{eqnarray}
\begin{array}{c}
\scalebox{.33}{\includegraphics*[trim=30 55 30 55,clip]{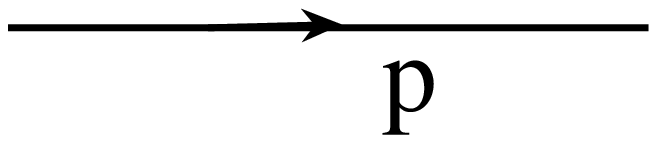}}
\end{array}\!\!\!\!
& = & \displaystyle{\frac {i}{2p^{+}p^{-} - \vec{p}^{\;2} - m^{2} +
  i\epsilon}}\nonumber\\
\noalign{\vskip -25pt}%
\begin{array}{c}
\scalebox{.33}{\includegraphics*{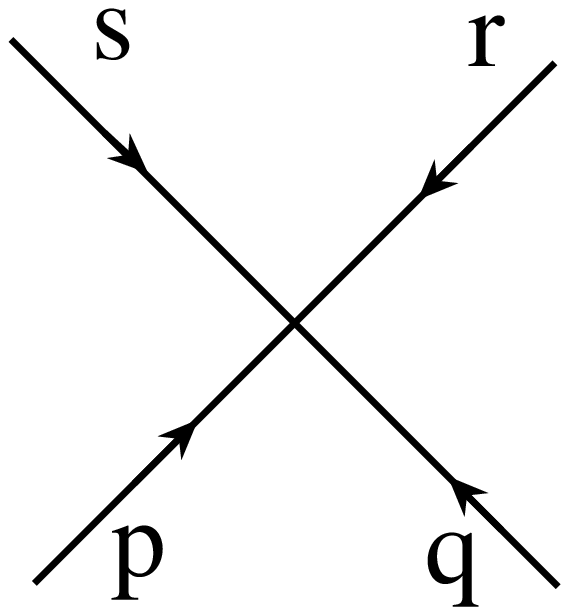}}
\end{array}\!\!\!\!
& = & \displaystyle{- i\lambda \delta^{n} (p+q+r+s)}
\end{eqnarray}
}
where all the momenta, at the vertex, are assumed to be incoming.

In going to finite temperature, as is well known, the interaction
vertices of the theory are unaffected, but the propagators modify
to reflect the periodicity (or anti-periodicity) of the field
variables \cite{kapusta,lebellac,das}. Let us generalize the theory in
eq. (\ref{lagrangian}) to 
finite temperature, following  the conventional identification
\begin{equation}
\langle {\cal O} \rangle_{\beta} = {\rm Tr}\;e^{-\beta H} {\cal O}
\equiv {\rm Tr}\;e^{-\beta P^{-}} {\cal O}\label{ensembleaverage}
\end{equation}
We can describe the resulting theory either in the real time
formalism or in the imaginary time formalism and we discuss the two
cases separately.

\subsection{Real time formalism}

Let us describe the propagators of the theory in the closed
time  path formalism \cite{schwinger,das} for simplicity. A similar
structure  for the propagators  results in
thermo field dynamics \cite{umezawa,das}, which we do not go into. It
is well known
that, in the real time formalism, the field degrees of freedom
double and the propagators have a $2\times 2$ matrix structure. In
the case of the light-front scalar field theory with the conventional
generalization in (\ref{ensembleaverage}), the propagators for the
doubled degrees of freedom have the forms
\begin{eqnarray}
iG_{++} (p) & = & \frac{i}{2p^{+}p^{-} - \omega_{p}^{2} +
i\epsilon} + 2\pi n_{B}(|p^{-}|)\, \delta
(2p^{+}p^{-}-\omega_{p}^{2})\nonumber\\
\noalign{\vskip 4pt}%
iG_{+-} (p) & = & 2\pi
\left(\theta (- p^{-}) + n_{B}(|p^{-}|)\right) \delta (2p^{+}p^{-}
- \omega_{p}^{2})\nonumber\\
\noalign{\vskip 4pt}%
iG_{-+} (p) & = & 2\pi \left(\theta
(p^{-}) + n_{B}(|p^{-}|)\right) \delta (2p^{+}p^{-} -
\omega_{p}^{2})\nonumber\\
\noalign{\vskip 4pt}%
iG_{--} (p) & = & -
\frac{i}{2p^{+}p^{-} - \omega_{p}^{2} - i\epsilon} + 2\pi n_{B}
(|p^{-}|)\, \delta (2p^{+}p^{-} -
\omega_{p}^{2})\label{naivepropagator}
\end{eqnarray}
where we have introduced the bosonic distribution function
\[
n_{B} (x) = \frac{1}{e^{\beta x} -1}
\]
and have defined 
\begin{equation}
\omega_{p}^{2} = \vec{p}^{\;2} + m^{2}
\end{equation}
It is worth remembering that $\omega_{p}$ involves only $(n-2)$
transverse components of the momenta, as opposed to the
conventional theories where it depends on all the $(n-1)$ spatial
components of the momentum. We also note that the $\pm$ subscripts
refer to the ``original" and the ``doubled" degrees of freedom
respectively. The propagators have the usual structure of a sum of
the zero temperature part and the finite temperature part.
However, the sign of trouble is already apparent in the form of
the propagators in (\ref{naivepropagator}). The thermal
distribution function does not seem to provide the necessary
damping, as is usual in conventional thermal field theories. Namely, for
$p^{-}=0$ (which is allowed by the delta function constraint), the
distribution function diverges. As we will see soon, more
difficulties arise in actual calculations.

\begin{figure}
\centerline{\includegraphics[width = 14cm, height = 9cm]{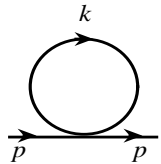}}
\vskip -4cm
\caption{One loop self-energy in $\phi^{4}$ theory}
\end{figure}

There are two kinds of vertices in the thermal field theory, ``$+$''
type  and
``$-$'' type, with a relative sign difference between the
two. However, at  one loop,
there is no mixing between the ``original" and the ``doubled"
degrees of freedom. As a result, the one loop correction to the
self-energy simply involves the tadpole graph (see Fig. 1), which can be
readily evaluated.
\begin{equation}
i \Pi_{++} (p) = - \frac{i\lambda}{2} \int \frac{d^{n}k}{(2\pi)^{n}}\;
iG_{++} (k)
\end{equation}
Separating out the zero temperature part, the thermal correction
to the self-energy can now be easily obtained. In fact, because of the
delta function, the $k^{-}$ integral can be trivially done leading
to
\begin{eqnarray}
i\Pi_{++}^{(\beta)} (p) & = & - \frac{i\lambda}{2(2\pi)^{n-1}}
\int d^{n}k\;n_{B}(|k^{-}|)\, \delta (2k^{+}k^{-} -
\omega_{k}^{2})\nonumber\\
\noalign{\vskip 4pt}%
 & = & - \frac{i\lambda}{2(2\pi)^{n-1}} \int d^{n-2}k \int_{0}^{\infty}
 \frac{dk^{+}}{k^{+}}\;n_{B}\left(\frac{\omega_{k}^{2}}{2k^{+}}\right)
\label{intermediate}
\end{eqnarray}
It is clear that the $k^{+}$ integral is divergent at the
ultraviolet limit, $k^{+}\rightarrow \infty$, and needs to be
regularized.  Regularizing the
power of $k^{+}$ in the denominator yields
\begin{eqnarray}
i\Pi_{++}^{(\beta)} (p) & = & \lim_{\epsilon\rightarrow
0}\;-\frac{i\lambda}{2(2\pi)^{n-1}} \int d^{n-2}k
\int_{0}^{\infty}
\frac{dk^{+}}{(k^{+})^{1+\epsilon}}\;n_{B}\left(\frac{\omega_{k}^{2}}{2k^{+}}
\right)\nonumber\\
\noalign{\vskip 4pt}%
 & = & \lim_{\epsilon\rightarrow 0}\;\frac{i\lambda}{4(2\pi)^{n-1}} 
\int d^{n-2}k
 \left(\frac{1}{\epsilon} - {\bf C} + \ln 2\pi + \ln
 \frac{2}{\beta\omega_{k}^{2}} + O(\epsilon)\right)\label{naivetadpole}
\end{eqnarray}

The form of the thermal correction in (\ref{naivetadpole}) is
exact and is  quite interesting. It shows that even though this
represents the thermal correction, it has a divergent part that is
independent of temperature. Thus, it would seem that in a thermal
background, the theory would require additional temperature
independent counterterms beyond the ones needed for the
regularization of the zero temperature theory. This is quite
distinct from the behavior of the conventional thermal field theories
and, if  true, would
cause enormous problems with the renormalizability properties of
the light-front theories at finite temperature. Furthermore, since
the integration over the $(n-2)$ transverse directions are yet to
be performed, we see that the temperature dependent part of the
amplitude is finite only in $1+1$ dimensions. In any other
dimension, the temperature dependent part diverges as well, requiring
temperature dependent counterterms. We would like to emphasize
here that, even though (\ref{naivetadpole}) represents the
temperature  dependent part of the one loop self-energy, it does
not vanish when temperature vanishes, namely as $\beta\rightarrow \infty$.
This is connected with the problem alluded to earlier, namely,
since $k^{-}$ can take a vanishing value, the limit
$\beta\rightarrow \infty$ of the distribution function is
ambiguous (it is non-analytic at that point). Further problems
arise when one studies the self-energy for the $\phi^{3}$ theory
on the light-front, but we will not go into the details of these here.
These are serious problems which suggest that the naive
generalization of ideas from conventional thermal field theories may not be
appropriate in the case of light-front field theories. In the next
section,  we will analyze the source of
these problems and propose the appropriate generalization for such
theories.

\subsection{Imaginary time formalism}

In the imaginary time formalism, we rotate the theory to Euclidean
space (imaginary time) and assume that the energy variable takes
discrete  values.
Consequently, the energy integrals are replaced by a sum over the
discrete Matsubara frequencies. In the light-front theories, this
translates to replacing, in the scalar propagator,
\begin{equation}
p^{-} \rightarrow 2i\pi n T
\end{equation}
where $T=\frac{1}{\beta}$ denotes temperature. As a result, the scalar
propagator, in the imaginary time formalism, becomes
\begin{equation}
G (p) =  \frac{1}{4i\pi nTp^{+} - \omega_{p}^{2}}
\end{equation}
The tadpole diagram, Fig. 1, is now straightforward to evaluate in the
imaginary time formalism,
\begin{eqnarray}
-\Pi (p) & = & -\frac{\lambda}{2} \int
\frac{d^{n-1}k}{(2\pi)^{n-1}}\; T\sum_{n=-\infty}^{\infty}
G(k)\nonumber\\
\noalign{\vskip 4pt}%
 & = & -\frac{\lambda}{2(2\pi)^{n-1}} \int d^{n-1}k\;T\sum_{n=-\infty}^{\infty}
 \frac{1}{4i\pi nTk^{+} - \omega_{k}^{2}}\nonumber\\
\noalign{\vskip 4pt}%
 & = & \frac{\lambda}{4(2\pi)^{n-1}} \int d^{n-2}k \int_{0}^{\infty}
 \frac{dk^{+}}{k^{+}}\; \coth
 \left(\frac{\omega_{k}^{2}}{4Tk^{+}}\right) 
\end{eqnarray}
Separating out the zero temperature contribution, we obtain,
\begin{equation}
-\Pi^{(\beta)} (p) =  \frac{\lambda}{4(2\pi)^{n-1}} \int d^{n-2}k
\int_{0}^{\infty} \frac{dk^{+}}{k^{+}} \left(\coth
\left(\frac{\omega_{k}^{2}}{4Tk^{+}}\right) -1\right) =
\frac{\lambda}{2(2\pi)^{n-1}} \int d^{n-2}k \int_{0}^{\infty}
\frac{dk^{+}}{k^{+}}\;n_{B}\left(\frac{\omega_{k}^{2}}{2k^{+}}\right)
\end{equation}
This is exactly the same expression as in (\ref{intermediate})
and, therefore, all the subsequent analysis of the earlier subsection
follows. The real time
and the imaginary time formalisms give the same result
which is, however, plagued by problems. We will discuss next the
source of the problem.

\section{The proper generalization to finite temperature}

To understand the source of the problems in the last section, let
us recapitulate briefly what happens in a conventional theory. In
a conventional thermal field theory, the thermal part of the
propagator represents the interactions of the particle with the
thermal distribution of real particles in the medium. This is
suppressed at high energies. In contrast, the propagators in eq.
(\ref{naivepropagator}), as we have argued, do not provide the
necessary damping. The form of the propagators are, of course, 
derived from the assumption that the ensemble average, in
light-front theories, is given by eq. (\ref{ensembleaverage}).
This is, in fact, where the problem lies. In a conventional
theory, when one assumes that the ensemble average has the form
\begin{equation}
\langle {\cal O} \rangle_{\beta} = {\rm Tr}\; e^{-\beta H}\; {\cal
O},\label{conventional}
\end{equation}
it is understood that we are in a Lorentz frame where the heat
bath is at rest. In fact, this is not a manifestly Lorentz
covariant description. One can give a manifestly covariant
description of thermal field theories \cite{israel,weldon} at the
expense  of
introducing a velocity for the heat bath, $u^{\mu}$, normalized to
unity, namely
\begin{equation}
u\cdot u = u^{\mu}u_{\mu} = 1\label{normalization}
\end{equation}
and generalizing the ensemble average to
\begin{equation}
\langle {\cal O} \rangle_{\beta} = {\rm Tr}\; e^{-\beta
u\cdot P}\, {\cal O} = {\rm Tr}\;e^{-\beta u^{\mu} P_{\mu}}\, {\cal O}
\label{covariant}
\end{equation}
In a conventional thermal field theory, where the metric is
diagonal and is of the form $(+,-,-,\cdots ,-)$, one can choose a rest
frame of the heat bath corresponding to $u^{\mu} =
(1,0,0,\cdots,0)$ consistent with (\ref{normalization}) and, in
this case, (\ref{covariant}) would reduce to the conventional
definition of ensemble average in (\ref{conventional}).

In contrast, a light-front description of a theory is manifestly
relativistic. Intuitively, it is clear that it is not possible to
have a heat bath at rest on the light-front. Eq.
(\ref{ensembleaverage}), on the other hand, is a generalization of
the rest frame ensemble average (\ref{conventional}) to light-front
theories  and,
consequently, there are bound to be problems. Note that, for
(\ref{covariant}) to reduce to (\ref{ensembleaverage}), we must
have $u^{\mu} = (1,0,0,\cdots ,0)$, which, with the light-front
metric, gives
\[
u^{\mu}u_{\mu} = 2 u^{+}u^{-} - \vec{u}\cdot \vec{u} = 0
\]
This is inconsistent with (\ref{normalization}) and this is
another way of saying that we cannot have a heat bath at rest on
the light-front.

It is clear, therefore, that in dealing with light-front field
theories, we must use a manifestly covariant description of the
thermal field theories. With this, let us now discuss 
the proper generalization of real time and imaginary time formalisms for
light-front field theories separately. In this section, we will restrict
ourselves to scalar field theories and describe fermion theories
in the next section.

\subsection{Real time formalism}

Let us assume that the heat bath is moving with a velocity
$u^{\mu}$ subject to (\ref{normalization}). In that case, in the
closed time path formalism, the propagators for the scalar field
theory take the forms (for the doubled degrees of freedom)
\begin{eqnarray}
iG_{++} (p) & = & \frac{i}{2p^{+}p^{-} - \omega_{p}^{2} +
i\epsilon} + 2\pi n_{B}(|u\cdot p|)\, \delta
(2p^{+}p^{-}-\omega_{p}^{2})\nonumber\\
\noalign{\vskip 4pt}%
iG_{+-} (p) & = & 2\pi
\left(\theta (- u\cdot p) + n_{B}(|u\cdot p|)\right) \delta
(2p^{+}p^{-} - \omega_{p}^{2})\nonumber\\
\noalign{\vskip 4pt}%
iG_{-+} (p) & = & 2\pi
\left(\theta (u\cdot p) + n_{B}(|u\cdot p|)\right) \delta
(2p^{+}p^{-} - \omega_{p}^{2})\nonumber\\
\noalign{\vskip 4pt}%
iG_{--} (p) & = & -
\frac{i}{2p^{+}p^{-} - \omega_{p}^{2} - i\epsilon} + 2\pi n_{B}
(|u\cdot p|)\, \delta (2p^{+}p^{-} -
\omega_{p}^{2})\label{propagator}
\end{eqnarray}
A simple choice for the velocity of the heat bath satisfying
(\ref{normalization}), for example, is (in the light-front basis)
\begin{equation}
u^{\mu} = \frac{1}{\sqrt{2}} (1,1,0,0,\cdots ,
0)\label{simplechoice}
\end{equation}
in which case, we have
\[
u\cdot p = \frac{1}{\sqrt{2}} (p^{+}+p^{-})
\]
and the bosonic distribution function takes the form
\[
n_{B}(|u\cdot p|) = n_{B} \left(\frac{1}{\sqrt{2}} |p^{+}+p^{-}|\right)
\]
It is easy to see that this distribution function provides the
necessary  damping, on-shell, both at $p^{+}=0$ and
$p^{+}\rightarrow \infty$.

With this modification of the propagators, we can now re-evaluate
the tadpole diagram (see Fig. 1). With (\ref{simplechoice}), the
temperature dependent part of the tadpole graph has the
form
\begin{eqnarray}
i\Pi_{++}^{(\beta)} (p) & = & -\frac{i\lambda}{2(2\pi)^{n-1}} \int
d^{n}k\;n_{B} \left(\frac{1}{\sqrt{2}} |k^{+}+k^{-}|\right) \delta
(2k^{+}k^{-} - \omega_{k}^{2})\nonumber\\
\noalign{\vskip 4pt}%
 & = & -\frac{i\lambda}{2(2\pi)^{n-1}} \int d^{n-2}k
 \int_{0}^{\infty} \frac{dk^{+}}{k^{+}}\;n_{B}
 \left(\frac{\omega_{k}^{2} + 2
 (k^{+})^{2}}{2\sqrt{2}k^{+}}\right)\label{behavior}
 \end{eqnarray}
 It is worth emphasizing that, unlike the corresponding expression
 with the naive generalization in (\ref{intermediate}), this
 integrand is well behaved in both the limits, $k^{+}=0$ and
 $k^{+}\rightarrow \infty$ as is expected of a thermal amplitude. As a
 result, it does not need any regularization. 
 Furthermore, it vanishes at zero temperature, $\beta\rightarrow
 \infty$, as we would expect since it represents the thermal
 correction to the self-energy. However, in
 general, it cannot be evaluated in a closed form. In the high
 temperature limit, $\beta m\ll 1$, and in four space-time dimensions,
 the integral has the value,
 \begin{equation}
 i\Pi_{++}^{(\beta)} (p) \approx -\frac{i\lambda}{24\beta^{2}} +
 O (\beta m)\label{tadpole}
 \end{equation}

 There are several things to note from this result. First of all,
 this yields a thermal mass correction which, in the high
 temperature limit, has the form
 \begin{equation}
 \Delta m_{\rm T}^{2} = \frac{\lambda}{24\beta^{2}} = \frac{\lambda
 T^{2}}{24} > 0
 \end{equation}
 Namely, the thermal mass correction is positive as is the case in
 conventional theories. This is, in fact, crucial for restoration
 of symmetry at finite temperature. More interestingly, we note
 that the thermal mass correction coincides exactly with that
 obtained in a conventional thermal (scalar) field theory in four
 dimensions. 

 Before closing this subsection on the real time formalism, let us
 note that the propagators (\ref{propagator}) satisfy the usual
 relations
 \begin{eqnarray}
 iG_{++} (p) + iG_{--} (p) & = & iG_{+-} (p) + iG_{-+}
 (p)\nonumber\\
\noalign{\vskip 4pt}%
 iG_{++} (p) - iG_{+-} (p) & = & iG_{-+} (p) - iG_{--} (p) =\;
 iG_{\rm R} (p)\; =\; \frac{i}{(u\cdot p)^{2} - (\bar{u}\cdot p)^{2} -
 \omega_{p}^{2} + i\epsilon\, {\rm sgn} (u\cdot p)}\nonumber\\
\noalign{\vskip 4pt}%
 iG_{++} (p) - iG_{-+} (p) & = & iG_{+-} (p) - iG_{--} (p) =\;
 iG_{\rm A} (p)\; =\; \frac{i}{(u\cdot p)^{2} - (\bar{u}\cdot p)^{2} -
 \omega_{p}^{2} - i\epsilon\, {\rm sgn} (u\cdot p)}
 \end{eqnarray}
 where $G_{\rm R},G_{\rm A}$ denote the retarded and the advanced
 propagators respectively and  we have defined a vector
 $\bar{u}^{\mu}$ which  is
 orthogonal to $u^{\mu}$ (as well as to any vector in the
 transverse $(n-2)$ dimensional space) and has a space-like normalization,
 namely,
 \begin{equation}
 u\cdot \bar{u} = 0,\qquad \bar{u}\cdot \bar{u} = -1
 \end{equation}
 For the choice in (\ref{simplechoice}), $\bar{u}^{\mu} =
 \frac{1}{\sqrt{2}} (1, -1,0,\cdots , 0)$.

 For completeness, let us also note here the forms of the
 propagator in the formalism of thermo field dynamics in
 light-front scalar field theories, with proper generalization.
 \begin{eqnarray}
 iG_{11} (p) & = & \frac{i}{2p^{+}p^{-} - \omega_{p}^{2} +
 i\epsilon} + 2\pi n_{B} (|u\cdot p|)\, \delta (2p^{+}p^{-} -
 \omega_{p}^{2})\nonumber\\
\noalign{\vskip 4pt}%
 iG_{12} (p) & = & 2\pi n_{B}(|u\cdot p|)\, e^{\frac{\beta
 |u\cdot p|}{2}} \;\delta (2p^{+}p^{-} -
 \omega_{p}^{2})\nonumber\\
\noalign{\vskip 4pt}%
 iG_{21} (p) & = & 2\pi n_{B} (|u\cdot p|)\, e^{\frac{\beta
 |u\cdot p|}{2}} \;\delta (2p^{+}p^{-} -
 \omega_{p}^{2})\nonumber\\
\noalign{\vskip 4pt}%
 iG_{22} (p) & = & - \frac{i}{2p^{+}p^{-} - \omega_{p}^{2} -
 i\epsilon} + 2\pi n_{B} (|u\cdot p|)\, \delta (2p^{+}p^{-} -
 \omega_{p}^{2})
 \end{eqnarray}

Finally, let us note that since, in this case, we have two
preferred vectors available, namely, $u^{\mu}$ and
$\bar{u}^{\mu}$, any given vector can be uniquely decomposed as
\begin{equation}
A^{\mu} = (A\cdot u) u^{\mu} - (A\cdot \bar{u}) \bar{u}^{\mu} +
A_{\rm T}^{\mu}\label{decomposition}
\end{equation}
where $A_{\rm T}^{\mu}$ is transverse to both $u,\bar{u}$.
Similarly, any higher order tensor structure can also be decomposed
with  respect
to these two vectors. In this way, a richer tensor structure
arises in light-front theories at finite temperature than in
conventional thermal field theories.

\subsection{Imaginary time formalism}

In the imaginary time formalism, in the covariant description, it
is the variable $(u\cdot p)$ that is rotated to Euclidean space and
takes discrete values. Thus,
\begin{equation}
u\cdot p \rightarrow  2i\pi nT
\end{equation}
which, with the choice of (\ref{simplechoice}), leads to
\begin{equation}
p^{-} \rightarrow 2\sqrt{2}i\pi nT - p^{+} = \bar{p}^{-} -
p^{+}\label{rotation}
\end{equation}
where we have identified
\begin{equation}
\bar{p}^{-} = 2\sqrt{2}i\pi nT
\end{equation}
(An alternate way of doing the rotation is to decompose the
momentum vector as in (\ref{decomposition}) and rotate $u\cdot p$
while treating $\bar{u}\cdot p$ as an independent variable.) As a result,
the scalar propagator, in the imaginary time formalism, takes the
form
\begin{equation}
G (p) = \frac{1}{2(\bar{p}^{-} - p^{+})p^{+} - \omega_{p}^{2}} =
\frac{1}{4\sqrt{2}i\pi nTp^{+} - (\omega_{p}^{2} +
2(p^{+})^{2})}\label{imaginarytimepropagator}
\end{equation}

With this, let us calculate the tadpole diagram, Fig. 1, in the $\phi^{4}$
theory.
\begin{eqnarray}
-\Pi (p) & = & - \frac{\lambda}{2(2\pi)^{n-1}} \int
d^{n-1}k\;\sqrt{2} T\sum_{n=-\infty}^{\infty} G(k,n)\nonumber\\
\noalign{\vskip 4pt}%
& = &
-\frac{\lambda}{2(2\pi)^{n-1}} \int d^{n-1}k\;\sqrt{2}T
\sum_{n=-\infty}^{\infty} \frac{1}{4\sqrt{2}i\pi nTk^{+} -
(\omega_{k}^{2} + 2(k^{+})^{2})}\nonumber\\
\noalign{\vskip 4pt}%
& = & \frac{\lambda}{4(2\pi)^{n-1}} \int d^{n-2}k
 \int_{0}^{\infty} \frac{dk^{+}}{k^{+}} \coth
 \left(\frac{\omega_{k}^{2} + 2(k^{+})^{2}}{4\sqrt{2}k^{+}T}\right)
 \end{eqnarray}
 Here, the factor of $\sqrt{2}$ arises from the Jacobian (because
 of the particular choice of the unit vector).
Separating out the zero temperature part, then, leads to the
thermal correction to the self-energy,
\begin{equation}
-\Pi^{(\beta)} (p) = \frac{\lambda}{2(2\pi)^{n-1}} \int d^{n-2}k
\int_{0}^{\infty} \frac{dk^{+}}{k^{+}}\;n_{B}
\left(\frac{(\omega_{k}^{2}+2(k^{+})^{2})}{2\sqrt{2}k^{+}}\right)
\end{equation}
which coincides with the result calculated earlier in the real
time formalism in (\ref{behavior}) and all the subsequent analysis
carries through.

\begin{figure}
\centerline{\includegraphics[width = 14cm, height = 9cm]{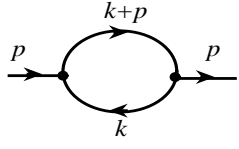}}
\vskip -4cm
\caption{One loop self-energy in $\phi^{3}$ theory}
\end{figure}

As another example, let us calculate the one loop scalar
self-energy in a massive $\phi^{3}$ theory in $3+1$ dimensions (see
Fig. 2). In the imaginary time formalism, this has the form
\begin{eqnarray}
-\Pi (p) & = & \frac{g^{2}}{2(2\pi)^{3}} \int d^{3}k\;\sqrt{2}T
\sum_{m} G(k,m) G(k+p,m)\nonumber\\
\noalign{\vskip 4pt}%
 & = & \frac{g^{2}}{2(2\pi)^{3}} \int d^{3}k\;\sqrt{2}T \sum_{m}
 \frac{1}{2(2\sqrt{2}i\pi mT - k^{+})k^{+} -
 \omega_{k}^{2}}\frac{1}{2(2\sqrt{2}i\pi mT + \bar{p}^{-} -
 k^{+}-p^{+})(k^{+}+p^{+}) - \omega_{k+p}^{2}}
\end{eqnarray}
The sum can be evaluated using standard formulae and leads to
\begin{eqnarray}
-\Pi (p) & = & - \frac{g^{2}}{16(2\pi)^{3}} \int
\frac{d^{3}k}{k^{+}(k^{+}+p^{+})}\,\frac{1}
{\frac{\omega_{k+p}^{2}+2(k^{+}+p^{+})^{2}}{2(k^{+}+p^{+})}
- \frac{\omega_{k}^{2}+2(k^{+})^{2}}{2k^{+}} - \bar{p}^{-}}\nonumber\\
\noalign{\vskip 6pt}%
 &  & \times \left(\coth
 \frac{\omega_{k}^{2}+2(k^{+})^{2}}{4\sqrt{2}k^{+}T} - \coth
 \left(\frac{\omega_{k+p}^{2}+2(k^{+}+p^{+})^{2}}{4\sqrt{2}(k^{+}+p^{+})T}
 - \frac{\bar{p}^{-}}{2\sqrt{2}T}\right)\right)
 \end{eqnarray}
 If we use here the fact that $\bar{p}^{-} = 2\sqrt{2}i\pi nT$
 as well as the periodicity of the hyperbolic function, the
 self-energy becomes (upon rotation to real time)
 \begin{eqnarray}
-\Pi (p) & = & - \frac{g^{2}}{16(2\pi)^{3}} \int
\frac{d^{3}k}{k^{+}(k^{+}+p^{+})}\,\frac{1}
{\frac{\omega_{k+p}^{2}+2(k^{+}+p^{+})^{2}}
{2(k^{+}+p^{+})} - \frac{\omega_{k}^{2}+2(k^{+})^{2}}{2k^{+}} -
(p^{-}+p^{+})}\nonumber\\
\noalign{\vskip 6pt}%
 &  & \times \left(\coth
 \frac{\omega_{k}^{2}+2(k^{+})^{2}}{4\sqrt{2}k^{+}T} - \coth
 \left(\frac{\omega_{k+p}^{2}+2(k^{+}+p^{+})^{2}}{4\sqrt{2}(k^{+}+p^{+})T}
 \right)\right)
 \end{eqnarray}

 It is now easy to take various limits of this expression. In
 fact, in the present case, we have more possibilities of taking
 limits than in a conventional thermal field theory
 \cite{weldon,weldon1}.  For example, we
 note  that if we set $p^{+} = \vec{p} = 0$
 and take the limit $p^{-}\rightarrow 0$, we obtain
 \begin{equation}
 -\Pi (p^{+}=0,p^{-}\rightarrow 0,\vec{p}=0) = 0
 \end{equation}
 On the other hand, if we set $p^{-}=\vec{p}=0$ and take the limit
 $p^{+}\rightarrow 0$, we obtain
 \begin{equation}
 -\Pi (p^{+}\rightarrow 0,p^{-}=0,\vec{p}=0) =
 \frac{g^{2}}{256\sqrt{2}\pi^{3}T} \int
 d^{3}k\;\left(\frac{1}{(k^{+})^{2}} - \frac{2}{\omega_{k}^{2}}\right)
 {\rm cosech}^{2}
 \frac{\omega_{k}^{2}+2(k^{+})^{2}}{4\sqrt{2}k^{+}T}
 \end{equation}
 Finally, we can also set $p^{-}=p^{+}=0$ and take the limit
 $\vec{p}\rightarrow 0$. In this limit, we obtain
 \begin{equation}
 -\Pi (p^{+}=0,p^{-}=0,\vec{p}\rightarrow 0) =
 \frac{g^{2}}{256\sqrt{2}\pi^{3}T} \int
 d^{3}k\;\frac{1}{(k^{+})^{2}} {\rm cosech}^{2}
 \frac{\omega_{k}^{2}+2(k^{+})^{2}}{4\sqrt{2}k^{+}T}
 \end{equation}
This shows that the three different ways of approaching the origin in
the energy-momentum space lead to quite different results. Thus,
light-front theories have a richer structure than the conventional
thermal field theories also in this sense. Note, however, that as
$T\rightarrow 0$, all the three limits lead to a vanishing result, as
would be expected in a zero temperature theory. Furthermore, an
interesting  question arises as to whether the three limits would lead to
newer definitions of masses in light-front theories (in a conventional
thermal  field theory,
we have only the screening mass and the plasmon mass corresponding to the
two possible limits that are allowed). Even
the question of what would correspond to the screening and the plasmon
masses in such theories remains an open question.

 \section{Fermion theories}

 The fermion theories, on the light-front are more tricky simply
 because the number of degrees of freedom decreases in this case. Let us
 consider, for example, a free massive fermion theory on the light-front
 described by the Lagrangian density
 \begin{equation}
 {\cal L} = \overline{\psi} (i\gamma^{\mu}\partial_{\mu} - m)\psi
 \end{equation}
 We can, of course, add interactions, but, as we know, interaction
 vertices are not modified at finite temperature. Therefore, it is
 sufficient to look at the free theory to determine the
 propagators at finite temperature.

 Let us define the light-front gamma matrices
 \begin{equation}
 \gamma^{\pm} = \frac{1}{\sqrt{2}} (\gamma^{0} \pm \gamma^{n-1})
 \end{equation}
 These are, in fact, nilpotent matrices, namely,
 \begin{equation}
 (\gamma^{\pm})^{2} = 0
 \end{equation}
 Defining the projection operators,
 \begin{equation}
 P^{(\pm)} = \frac{1}{2} \gamma^{\mp}\gamma^{\pm} = \frac{1}{2}
 (1\pm \alpha_{n-1})
 \end{equation}
where $\vec{\alpha}$ represents the Dirac matrices, it is easy to
check that
\begin{equation}
(P^{(\pm)})^{2} = P^{(\pm)},\qquad P^{(+)}P^{(-)} = 0 =
P^{(-)}P^{(+)},\qquad P^{(+)} + P^{(-)} = 1
\end{equation}

With these projection operators, let us define
\begin{equation}
\psi^{(\pm)} = P^{(\pm)} \psi
\end{equation}
Then, it follows from the properties of the gamma matrices that
\begin{equation}
\gamma^{-}\psi^{(+)} = 0 = \gamma^{+}\psi^{(-)}
\end{equation}
This allows us to write the Lagrangian density in the light-front
variables as
\begin{eqnarray}
{\cal L} & = & \sqrt{2}\left[\psi^{(+)^{\dagger}} i\partial_{+}\psi^{(+)}
+ \psi^{(-)^{\dagger}} i\partial_{-}\psi^{(-)}\right.\nonumber\\
\noalign{\vskip 4pt}%
 &  & \qquad \left. -\frac{1}{2} \psi^{(+)^{\dagger}}
 \gamma^{-}(i\vec{\gamma}\cdot \vec{\nabla} + m)\psi^{(-)} -
 \frac{1}{2} \psi^{(-)^{\dagger}} \gamma^{+}(i\vec{\gamma}\cdot
 \vec{\nabla} + m)\psi^{(+)}\right]
 \end{eqnarray}
 It is clear now that only the $\psi^{(+)},\psi^{(+)^{\dagger}}$
 degrees of freedom are dynamical. The other degrees of freedom
 are related to these and can be eliminated.

 The fermion propagator, at zero temperature, has been derived
 long ago \cite{chang2} and has the form
 \begin{equation}
 iS_{F}(x-y) = \langle 0|T^{+}(\psi(x)\overline{\psi}(y))|0\rangle =
 \sqrt{2}i \int \frac{d^{n}p}{(2\pi)^{n}} e^{-ip\cdot (x-y)}
 \left(\frac{p\!\!\!\slash + m}{p^{2}-m^{2}+i\epsilon} -
 \frac{\gamma^{+}}{2p^{+}}\right)
 \end{equation}
where ``$T^{+}$" denotes ordering with respect to $x^{+}$. Thus,
we can identify
 \begin{equation}
 S_{F} (p) = \sqrt{2}\left(\frac{p\!\!\!\slash + m}{p^{2}-m^{2}+i\epsilon} -
 \frac{\gamma^{+}}{2p^{+}}\right) =
 \sqrt{2}\;\frac{\bar{p}\!\!\!\slash +
 m}{p^{2}-m^{2}+i\epsilon}\label{complete}
 \end{equation}
 where $\bar{p}^{\mu}$ denotes an on-shell momentum, namely,
 \begin{equation}
 \bar{p}^{\, +}=p^{+},\qquad \vec{\bar{p}}=\vec{p},\qquad \bar{p}^{\, -}
 = \frac{\vec{p}^{2}+m^{2}}{2p^{+}}
 \end{equation}
 The second form of the propagator makes it very clear that, when
 properly normalized, it can be thought of as a projection
 operator and, consequently, its inverse does not exist. This is,
 therefore, not a suitable structure to generalize to finite
 temperature. On the other hand, the singular structure of the
 zero temperature propagator simply reflects the fact that there are
 constraints in the theory, namely, that not all the degrees of
 freedom are dynamical. If we eliminate the non-dynamical degrees
 of freedom, the entire theory can be recast in terms of
 $\psi^{(+)},\psi^{(+)^{\dagger}}$ variables. Therefore, the
 relevant propagator, from the point of view of the theory, is
 \begin{equation}
 \overline{S}_{F} (x-y) = \langle 0|T^{+}(\psi^{(+)}(x)
 \psi^{(+)^{\dagger}}(y))|0\rangle = \langle 0|P^{(+)}
 T^{+}(\psi(x)\overline{\psi}(y)) \gamma^{0}P^{(+)}|0\rangle
 = \int
 \frac{d^{n}p}{(2\pi)^{n}}\;e^{-ip\cdot(x-y)}\,\overline{S}_{F} (p)
 \end{equation}
This can, in fact, be calculated from (\ref{complete}) in a simple
manner and is determined to be
\begin{equation}
\overline{S}_{F} (p) = \frac{\sqrt{2}p^{+}}{p^{2}-m^{2}+i\epsilon} =
\frac{\sqrt{2}p^{+}}{2p^{+}p^{-}-\omega_{p}^{2}+i\epsilon}
\end{equation}
This is well behaved with the two point function, in this
projected space, corresponding to
\[
\frac{p^{2}-m^{2}}{\sqrt{2}p^{+}} =
\frac{2p^{+}p^{-}-\omega_{p}^{2}}{\sqrt{2}p^{+}} 
\]
and this propagator can be easily generalized to finite
temperature. We note that this form of the propagator has also been
calculated earlier directly from the field decomposition in
\cite{prem}. 

The fermion propagators at finite temperature, in the real time
formalism  (closed time path), now take the forms
\begin{eqnarray}
i\overline{S}_{++} (p) & = &
\sqrt{2}p^{+}\left(\frac{i}{2p^{+}p^{-}-\omega_{p}^{2}+i\epsilon} - 2\pi
n_{F}(|u\cdot p|)\, \delta
(2p^{+}p^{-}-\omega_{p}^{2})\right)\nonumber\\
\noalign{\vskip 4pt}%
i\overline{S}_{+-} (p) & = & -2\sqrt{2}\pi p^{+}
\left(n_{F}(|u\cdot p|) - \theta (-u\cdot p)\right)\delta
(2p^{+}p^{-}-\omega_{p}^{2})\nonumber\\
\noalign{\vskip 4pt}%
i\overline{S}_{-+} (p) & = &-2\sqrt{2}\pi
p^{+}\left(n_{F}(|u\cdot p|) - \theta(u\cdot p)\right)\delta
(2p^{+}p^{-}-\omega_{p}^{2})\nonumber\\
\noalign{\vskip 4pt}%
i\overline{S}_{--} (p) & = &
\sqrt{2}p^{+}\left(-\frac{i}{2p^{+}p^{-}-\omega_{p}^{2}-i\epsilon} -2\pi
n_{F}(|u\cdot p|)\, \delta (2p^{+}p^{-}-\omega_{p}^{2})\right)
\end{eqnarray}
where $n_{F}$ represents the fermion distribution function
\[
n_{F}(x) = \frac{1}{e^{\beta x} + 1}
\]
The propagators in thermo field dynamics can similarly be obtained
and we do not go into this here.
In the imaginary time formalism, the fermion propagator, for the
independent degrees of freedom, takes the form
\begin{equation}
\overline{S} (p) = \frac{\sqrt{2}p^{+}}{2\sqrt{2}i(2n+1)\pi p^{+}T
- (\omega_{p}^{2}+ 2(p^{+})^{2})}
\end{equation}

\section{Summary}

In this paper, we have described how light-front field theories can be
generalized to finite temperature. We have shown that the naive
generalization leads to problems and the origin of the difficulty is
identified. Since light-front field theories describe relativistic
systems, a covariant description of thermal field theories becomes
necessary for the proper formulation of thermal light-front
theories. We discuss scalar and fermion light-front field theories at
finite temperature in detail, including issues such as non-analyticity
of self-energy and tensor decomposition. Several open questions are
also discussed. Light-front gauge theories at
finite temperature as well as further applications are presently under
study and will be reported later.  
\vskip .5cm
\noindent{\bf Acknowledgment:}

This work was supported in part by US DOE
Grant number DE-FG 02-91ER40685 and by CAPES, Brasil.

\end{document}